\documentstyle[12pt]{article}

\def\noi{\noindent}
\def\lam{\lambda}

\def\del{\partial}
\def\nab{\nabla}
\def\til{\tilde}
\def\dis{\displaystyle}

\def\a{\rm a}
\def\b{\rm b}

\def\A{\rm A}

\begin{document}

\vspace*{.7cm}

\begin{center}
{\large\bf On the equivalence theorem in $f(R)$-type generalized gravity}
\\[10mm]
\end{center}

\hspace*{15mm}\begin{minipage}{13.5cm}
Y. Ezawa, H. Iwasaki, Y. Ohkuwa$^{\dagger}$, S. Watanabe, N. Yamada and 
T. Yano$^{*}$\\[3mm]
Department of Physics, Ehime University, Matsuyama, 790-8577, Japan\\[1mm]
\hspace*{-1.5mm}$^{\dagger}$Section of Mathematical Science, Department of 
Social Medicine, Faculty of Medicine, University of Miyazaki, Kiyotake, 
Miyazaki, 889-1692, Japan\\[1mm]
\hspace*{-1.5mm}$^{*}$Department of Electrical Engineering, Ehime University, 
Matsuyama, 790-
8577, Japan\\[5mm]
{\small Email :  ezawa@sci.ehime-u.ac.jp, hirofumi@phys.sci.ehime-u.ac.jp,\\
ohkuwa@med.miyazaki-u.ac.jp, shizuka@phys.sci.ehime-u.ac.jp,\\
naohito@phys.sci.ehime-u.ac.jp and yanota@eng.ehime-u.ac.jp}
\\[8mm]
{\bf Abstract}

We investigate whether the equivalence theorem in $f(R)$-type gravity is valid
also in quantum theory.
It is shown that, if the canonical quantization is assumed, the equivalence 
does not hold in quantum theory.
\end{minipage}

\section{Introduction}

Recently, generalizations of Einstein gravity, or, higher curvature gravity 
theories,  receive much attention.
Most of them are devided into two classes.
One of them is the Lovelock theory-type one in which the generalizations are 
made under the assumption that the equations of motion are 2nd order 
differential equations\cite{Lov}.
This class includes the Einstein gravity corrected by the Gauss-Bonnet terms, 
which is also motivated by string theory since the construction of the model 
relys on perturbation method.
Another is the so called fourth order gravity in which the equations of motion 
are fourth order differential equations.
The $f(R)$-type gravity is the typical one of this class\cite{C-DF}.

If the the true theory would turn out to belong to the former class , that 
would be desirable since structure of the theory is simpler.
However, present status is far from definitive, so investigations of the 
latter class of theories are attracting much attention.

In this work, we investigate the structure of the $f(R)$-type gravity, 
especially the conformal equivalence of the theory with the Einstein gravity 
coupled to a scalar field.
Classically both theories give the same solution corresponding to the path of 
stationary action\cite{TT,Whit,BC,Maeda,Wands,MS,FG,FN}.
However quantum fluctuations might invalidate the equivalence.
The fluctuations should be induced by contributions of various paths in the 
path integral formalism or fundamental commutation relations, although their 
precise estimations are not known yet.

In canonical quantum theory, commutation relations among fundamental canonical 
variables are related to the corresponding Poisson brackets in the classical 
theory.
Thus if the equivalence holds also in the quantum theory, the fundamental 
Poisson brackets should be equivalent in both theories, i.e. Poisson brackets 
of the conformally transformed theory should be the same as calculated using 
the original Poisson brackets.
So we examine whether this is the case or not for $f(R)$-type gravity. That is,
whether the conformal transformation, the coordinate transformation in the 
phase space, is a canonical transformation or not.

In section 2, transformation of canonical variables corresponding to the 
conformal transformation is carried out and express the canonical variables 
after the transformation as functions of the original canonical variables 
using the canonical formalism of \cite{EIOWYY}.
The transformation is not the point transformation in the formalism of 
$f(R)$-type gravity.
In section 3, we examine the equivalence of the two sets of Poisson brackets 
of canonical variable before and after the transformation using the results of 
section 2.
It is shown that two sets of Poisson brakets are not compatible.
Section 4 is devoted to the summary and discussion.

\section{Conformal transformation in terms of canonical\\ variables}

\subsection{Canonical variables}

We start from the action in the Jordan frame
$$
S=\int d^Dx\,{\cal L}=\int d^Dx\sqrt{-g}f(R)                        \eqno(2.1)
$$
where $D$ is the dimension of multidimensional spacetime, $g$ the determinant 
of multidimensional metric $g_{\mu\nu}$ and $R$ the multidimensional scalar 
curvature.
$f(R)$ is an almost arbitrary function of $R$, typical form of which is 
expressed as
$$
f(R)=R^{-s}\sum_{k=0}a_{k}R^k.                                     \eqno(2.2)
$$
$s$ is usually taken to be 1.
Of course, $k=s$ term is the cosmological constant term and $k=s+1$ term is 
the Einstein gravity term.
Field equations derived from (2.1) are expressed in the following form
$$
G_{\mu\nu}={1\over f'(R)}\left[\nab_{\mu}\nab_{\nu}f'(R)-g_{\mu\nu}\Box f'(R)
+{1\over2}g_{\mu\nu}\Bigl\{f(R)-Rf'(R)\Bigr\}\right]              \eqno(2.3\a)
$$
where $G_{\mu\nu}$ is the Einstein tensor and $f'(R)\equiv df/dR$.
Taking the trace of (2.3a), we have
$$
{1\over f'}\Box f'-{d+1\over 2d}{f(R)\over f'(R)}+{1\over d}R=0.  \eqno(2.3\b)
$$

Canonical formalism in \cite{EIOWYY} is the generalization of Ostrogradski's 
formalism\cite{Ost} by taking the advantageous point of the one by Buchbinder 
and Lyakhovich\cite{BL}.
That is to say, in defining the new generalized coordinate, time derivatives 
used in the former formalism are replaced by Lie derivatives along a timelike 
vector.
Components of the metric are decomposed following ADM:
$$
g_{\mu\nu}=\left(\begin{array}{cc}
-N^2+N_{k}N^k&N_{i}\\[3mm]
N_{i}&\ h_{ij}\ 
\end{array}\right)                                                 \eqno(2.4)
$$
where $i$ and $j$ run from 1 to d with $D=1+d$.

As generalized coordinates, we take $h_{ij},N,N^i$ and $K_{ij}$ which is 
(twice) the Lie derivative of $h_{ij}$, or the extrinsic curvature of 
$\Sigma_{t}$  and is known to be given as
$$
K_{ij}={1\over 2N}\left(\del_{0}h_{ij}-N_{i;j}-N_{j;i}\right),     \eqno(2.5)
$$
which will be denoted as $Q_{ij}$ hereafter.
Canonically conjugate momenta to $N$ and $N^i$ vanish and those to $h_{ij}$ 
and $Q_{ij}$ are denoted as $p^{ij}$ and $P^{ij}$ and are given as\cite{EIOWYY}
$$
p^{ij}=-\sqrt{h}\left[f'(R)Q^{ij}+h^{ij}f''(R){\cal L}_{n}R\right] \eqno(2.6\a)
$$
and
$$
P^{ij}=2\sqrt{h}f'(R)h^{ij}.                                      \eqno(2.6\b)
$$
The Lie derivatives are calculated along the normal vector $n=N^{-1}(1,N^i)$ 
to $\Sigma_{t}$.
From equation (2.6b), $f'(\hat{R})$ is expressed as follows
$$
f'(R)={h_{kl}P^{kl}\over 2d\sqrt{h}}={P\over 2d\sqrt{h}}          \eqno(2.7\a)
$$
where $P\equiv h_{ij}P^{ij}$, the trace of $P^{ij}$. 
Therefore $P^{ij}$ has only the trace part which is expressed by $P$.
Solving (2.7a) for $R$, we denote the solution as
$$
R=f'^{-1}\left({P\over 2d\sqrt{h}}\right)\equiv 
\Psi\left({P\over 2d\sqrt{h}}\right).                              \eqno(2.8)
$$
Correspondingly, traceless part of $Q^{ij}$(denoted as $Q^{\dagger\;ij}$) 
is given by (2.6a) as
$$
Q^{\dagger \;ij}=-{1\over \sqrt{h}f'(R)}p^{\dagger\, ij}
=-{2d\over P}\,p^{\dagger\,ij}                                    \eqno(2.9\a)
$$
or
$$
Q^{ij}=-{2d\over P}p^{\dagger\,ij}+{1\over d}h^{ij}Q.             \eqno(2.9\b)
$$
The generalized coordinate conjugate to $P$ would be $Q\equiv h^{ij}Q_{ij}$, 
the trace of $Q_{ij}$.
However, the Poisson bracket between $Q$ and $P$ is
$$
\{Q,P\}=d.                                                         \eqno(2.10)
$$
Thus the canonical pair is $\dis \Bigl(Q,{1\over d}P\Bigr)$ which we denote as 
$(Q,\Pi)$, i.e. 
$$
\Pi\equiv {1\over d}P.                                             \eqno(2.7\b)
$$

In terms of these variables scalar curvature $R$ is expressed as
$$
R=2N^{-1}\del_{0}Q+\left({2\over \Pi}\right)^2p^{\dagger\,ij}
p^{\dagger}_{\;ij}
+{d+1\over d}Q^2+{}^d\!R-2N^{-1}N^k\del_{k}Q-2N^{-1}\Delta N       \eqno(2.11)
$$
where ${}^d\!R$ is the scalar curvature of $\Sigma_{t}$.
The velocity $\del_{0}Q$ is expressed in terms of the generalized coordinates 
and their canonical momenta as
$$
\del_{0}Q={1\over2}N\Bigl[\Psi(\Pi/2\sqrt{h})-4\Pi^{-2}\Bigl(p^{ij}p_{ij}
-{1\over d}p^2\Bigr)-{d+1\over d}Q^2-{}^d\!R\Bigr]+N^k\del_{k}Q+\Delta N.
                                                                   \eqno(2.12)
$$

\subsection{Conformal transformation}

It is well known that the following conformal transformation makes the 
$f(R)$-type gravity to Einstein gravity with a scalar field so the transformed 
frame is referred to as Einstein frame:
$$
\til{g}_{\mu\nu}=\left[2\kappa^2f'(R)\right]^{2/(d-1)}{g}_{\mu\nu},\ \ \ 
\kappa^2=8\pi G                                                    \eqno(2.13)
$$
The scalar field $\phi$ is defined as
$$
\kappa\phi\equiv \sqrt{d/(d-1)}\ln[2\kappa^2f'(R)],\ \ \ {\rm or}\ \ \ 
f'(R)={1\over2}\kappa^{-2}\exp{\left(\kappa\sqrt{(d-1)/d}\,\phi\right)}.      
                                                                  \eqno(2.14\a)
$$
Solving for $R$, we have
$$
R=f'^{-1}\left({1\over2}\kappa^{-2}\exp{\Bigl\{\kappa\sqrt{(d-1)/d}\phi\Bigr\}}
\right)\equiv r(\phi).                                            \eqno(2.14\b)
$$
Field equations in the transformed frame are written as
$$
\til{G}_{\mu\nu}=\kappa^2\left[\del_{\mu}\phi\,\del_{\nu}\phi
-{1\over2}\til{g}_{\mu\nu}
\left\{\til{g}^{\lam\rho}\del_{\lam}\phi\,\del_{\rho}\phi
-\kappa^{-2}\exp{\Bigl\{-2\kappa\phi/\sqrt{d(d-1)}\Bigr\}}\Bigl({f\over\,f'\,}
-r\Bigr)\right\}\right].                                           \eqno(2.15)
$$
This is the Einstein equation for the transformed metric $\til{g}_{\mu\nu}$
with the scalar field $\phi$ as the source.
Eq.(2.15) is derived from the action
$$
\til{S}=\int\sqrt{-\til{g}}({\cal L}_{G}+{\cal L}_{M})d^Dx         \eqno(2.16)
$$
with
$$
{\cal L}_{G}={1\over 2\kappa^2}\til{R},\ \ \ 
{\cal L}_{M}
=-{1\over2}\til{g}^{\lam\rho}\del_{\lam}\phi\,\del_{\rho}\phi-V(\phi)
                                                                   \eqno(2.17)
$$
where
$$
V(\phi)=\exp{\Bigl\{-\kappa(d+1)/\sqrt{d(d-1)}\phi\Bigr\}}\left[
{\kappa^{-2}\over2}\exp{\Bigr\{\kappa\sqrt{(d-1)/d}\phi\Bigr\}}r(\phi)
-f(r(\phi))\right].                                                \eqno(2.18)
$$
Eq.(2.18) comes from the fact that the coefficient of $\til{g}_{\mu\nu}$ in 
(2.15) is ${\cal L}_{M}$.
In the following, we use a unit for which $2\kappa^2=1$ for simplicity.

Conformal transformations of the ADM variables are expressed as follows:
$$
\til{h}_{ij}=\Bigl[f'(R)\Bigr]^{2/(d-1)}h_{ij},\ \ \ \til{N}^i=N^i\ \ \ 
{\rm and}\ \ \ \til{N}=\Bigl[f'(R)\Bigr]^{1/(d-1)}N.             \eqno(2.19\a)
$$
The following transformations are also useful,
$$
\til{h}^{ij}=\Bigl[f'(R)\Bigr]^{-2/(d-1)}h^{ij},\ \ \ 
\til{N}_{i}=\Bigl[f'(R)\Bigr]^{2/(d-1)}N_{i},\ \ \ 
\til{h}=\Bigl[f'(R)\Bigr]^{2d/(d-1)}h.                           \eqno(2.20\a)
$$
In terms of the canonical variables, equations (2.19a) are expressed as 
follows:
$$
\til{h}_{ij}=\biggl[{\Pi\over 2\sqrt{h}}\biggr]^{2/(d-1)}h_{ij},\ \ \ 
\til{N}^i=N^i\ \ \ {\rm and}\ \ \ 
\til{N}=\biggl[{\Pi\over 2\sqrt{h}}\biggr]^{1/(d-1)}N             \eqno(2.19\b)
$$
where (2.7b) is used.
The scalar field $\phi$ is expressed as
$$
\phi=\sqrt{2d/(d-1)}\,\ln\Bigl[{\Pi\over 2\sqrt{h}}\Bigr].         \eqno(2.21)
$$
Similarly, (2.20a) are rewitten as
$$
\til{h}^{ij}=\biggl[{\Pi\over 2\sqrt{h}}\biggr]^{-2/(d-1)}h^{ij},\ \ \ 
\til{N}_{i}=\biggl[{\Pi\over 2\sqrt{h}}\biggr]^{2/(d-1)}N_{i}\ \ \ 
{\rm and}\ \ \ \til{h}=\biggl[{\Pi\over 2\sqrt{h}}\biggr]^{2d/(d-1)}h.
                                                                  \eqno(2.20\b)
$$

As in the Jordan frame, momenta canonically conjugate to $\til{N}$ and 
$\til{N}^i$ vanish and those conjugate to $\til{h}_{ij}$ and $\phi$, denoted 
as $\til{p}^{ij}$ and $\pi$ respectively, are given as
$$
\hspace{-1mm}\left\{
\begin{array}{ll}
\til{p}^{ij}&\!\!
             =\sqrt{\til{h}}\,\Bigl[\til{K}_{ij}-\til{h}^{ij}\til{K}\Bigr]
\\[3mm]
         &\!\!=\dis \Bigl[{\Pi\over 2\sqrt{h}}\Bigr]^{(d-3)/(d-1)}\sqrt{h}
              \left[-{2\over \Pi}\Bigl(p^{ij}-{1\over d}h^{ij}p\Bigr)
               +h^{ij}\Bigl\{{1\over d}Q+N^{-1}N^k_{\;;k}
              -(N\Pi)^{-1}(\del_{0}\Pi-N^k\Pi_{;k})\Bigr\}\right]
\\[7mm]
\pi&\!\!=-\sqrt{-\til{g}}\,\til{g}^{0\mu}\del_{\mu}\phi
          =\til{N}^{-1}\sqrt{\til{h}}\,(\del_{0}\phi-\til{N}^i\del_{i}\phi)
\\[3mm]
&\!\!\dis
      =\sqrt{d/2(d-1)}\,N^{-1}\Bigl[\del_{0}\Pi
       -N^i\Pi_{;i}-\Pi(NQ+N^k_{\ ;k})\Bigr]
\end{array}\right.                                                 \eqno(2.22)
$$
where we used a relation
$$
\til{K}_{ij}=\Bigl[{P\over 2d\sqrt{h}}\Bigr]^{1/(d-1)}\Bigl[Q_{ij}
+{1\over d-1}(NP)^{-1}h_{ij}\Bigl\{\del_{0}P-N^kP_{;k}
            -P(NQ+N^k_{\;k})\Bigr\}\Bigr].                         \eqno(2.23)
$$

\section{Poisson brackets}

In this section, we examine whether the theories before and after the 
conformal transformation are equivalent quantum mechanically.
As mentioned in the introduction, we adress this problem by examining two sets 
of Poisson brackets(PBs) among the fundamental canonical variables.
If we assume the canonical quantization, commutation relations among the 
canonical variables are proportional to corresponding Poisson brackets.
So quantum mechanical equivalence of two theories, would require that two sets 
of the PBs should be equivalent.
The canonical variables in the Einstein frame are functions of the ones in the 
Jordan frame as are given by (2.19b), (2.21) and (2.22).
These are not point transformations but rather complicated coordinate 
transformations of the phase space, so that the equivalence of PBs are not 
evident.
PBs among the former variables, however, can be calculeted in terms of the 
latters and the consistency can be checked.

The fundamental canonical variables in the original $f(R)$-type theory 
are $(h_{ij},\ Q,\ N,\ N^i;\ $\\
$p^{ij},\ \Pi,\ p_{N},\ p_{i})$ where $p_{N}$ and $p_{i}$ are constrained to 
vanish and their nonvanishing PBs are expressed as follows:
$$
\{h_{ij}({\bf x},t),\ p^{kl}({\bf y},t)\}
={1\over2}(\delta_{i}^k\delta_{j}^l+\delta_{i}^l\delta_{j}^k)
\delta({\bf x}-{\bf y})\ \ {\rm and}\ \ 
\{Q({\bf x},t),\Pi({\bf y},t)\}=\delta({\bf x}-{\bf y}).           \eqno(3.1)
$$
Similarly, in the Einstein frame, nonvanishing fundamental PBs are expressed
as follows:
$$
\{\til{h}_{ij}({\bf x},t),\ \til{p}^{kl}({\bf y},t)\}
={1\over2}(\delta_{i}^k\delta_{j}^l+\delta_{i}^l\delta_{j}^k)
\delta({\bf x}-{\bf y})\ \ {\rm and}\ \ 
\{\phi({\bf x},t),\pi({\bf y},t)\}=\delta({\bf x}-{\bf y}).        \eqno(3.2)
$$
PBs in (3.2) should be derivable using (3.1).
This could in principle be carried out straightfowardly by taking the tilde 
quantities as functions of original canonical variables.
For the generalized coordinates, PBs among them are easily calculated from 
(2.19b) and (2.21) and we have
$$
\begin{array}{l}
\{\til{h}_{ij}({\bf x},t),\til{h}_{kl}({\bf x},t)\}
=\{\til{h}_{ij}({\bf x},t),\til{N}({\bf y},t)\}
=\{\til{h}_{ij}({\bf x},t),\til{N}^k({\bf y},t)\}
=\{\til{N}({\bf x},t),\til{N}^i({\bf y},t)\}\\[3mm]
=\{\til{h}_{ij}({\bf x},t),\phi({\bf y},t)\}
=\{\til{N}({\bf x},t),\phi({\bf y},t)\}
=\{\til{N}^i({\bf x},t),\phi({\bf y},t)\}=0.
\end{array}                                                        \eqno(3.3)
$$
However, for PBs involving the canonical momenta, the calculations are lengthy 
and complex partly because the time derivative of $\Pi$, the momentum 
canonically conjugate to $Q$, is an arbitrary function before we use the 
equation of motion, so has to be determined from the consistency of (3.1) and 
(3.2) expressed by a set of partial differential equations.
These equations are complicated.
Thus instead of solving these equations, we will show that a contradiction 
arises if we assume both of (3.1) and (3.2) to hold.

From $\{\til{h}_{ij}({\bf x},t),\pi({\bf y},t)\}=0$, we have
$$
\{\til{h}_{ij}({\bf x},t),\del_{0}\Pi({\bf y},t)\}
=N\Pi({\bf y},t)\{\til{h}_{ij}({\bf x},t),Q({\bf y},t)\}.          \eqno(3.4)
$$ 
Using (3.4) and expressing the transformed variables in terms of the original
variables, we have
$$
\{\til{h}_{ij}({\bf x},t),\til{p}^{kl}({\bf y},t)\}
=-\left[{1\over2}(\delta_{i}^k\delta_{j}^l+\delta_{i}^l\delta_{j}^k)
-{2\over d}\til{h}_{ij}\til{h}^{kl}\right]\delta({\bf x}-{\bf y})  \eqno(3.5)
$$
which contradicts (3.2).

\section{Summary and discussions}

We investigated whether the equivalence theorem in $f(R)$-type gravity holds
in the quantum theoretical level.
If we assume the canonical quatization, commutation relations of funadamental
variables are proportional to the corresponding Poisson brackets.
Therefore, if the equivalence remains valid also in quantum theory, 
equivalence of the fundamental Poisson brackets in Jordan frame and Einstein 
frame should be necessary.
We examined this necessary condition and showed that it does not hold.
Therefore quantum equivalence of both frames would not hold.

However, if we would quantize noncanonically\cite{NCQ,KJS}, e.g. in terms of 
noncommutative 
geometry\cite{NCG,Wess}, there would be a possibility of recovering the 
equivalence.
In other words, if we introduce noncommutativity through Poisson brackets, 
according to ref.\cite{KJS}, it would be possible that the $f(R)$-type generalized 
gravity is equivalent to Einstein gravity with a scalar field in which 
dynamical variables are noncommutative.
This would be, in a sense, natural since both higher curvature effects and 
noncommutativity would appear at short distances. 
Investigation of this possibility would be interesting.

Finally we comment on the relation to classical equivalence.
Classically, the variational principle is imposed and the path is chosen to 
make the action stationary.
In the calculation of PBs, non-minimum paths are taken into account.
The violation of the equivalence could be interpreted to arise from the 
contribution of these paths.
In this sense, quantum non-equivalence is similar to the quntum anomaly.
\\[3mm]

\noi
{\Large\bf Appendix : Examples of deformed Poisson brackets}\\[3mm]
In this apendix, we present the PBs among the fundamental variables in 
Einstein frame in terms of Jordan frame variables and provide examples of 
deformed PBs under some simplifying assumptions.\\

\noi
{\large\bf A1. Poisson brackets among Einstein frame variables}\\[3mm]
PBs among the fundamental variables of Einstein frame variables are calculated 
from those of Jordan frame variables as follows:
$$
\begin{array}{l}
\dis \{\til{h}_{ij}({\bf x},t),\pi({\bf y},t)\}={1\over 2d}f^3F^2
\Bigl[h_{ij}\Bigl(A+\delta({\bf x}-{\bf y})\Bigr)+{d-1\over 2}C_{ij}\Bigr]
\\[3mm]
\dis \{\til{h}_{ij}({\bf x},t),\til{p}^{kl}({\bf y},t)\}
=-\Bigl[{1\over2}(\delta_{i}^k\delta_{j}^l+\delta_{i}^l\delta_{j}^k)
-{d-2\over d(d-1)}h_{ij}h^{kl}\Bigr]\delta({\bf x}-{\bf y})
\\[4mm]
\dis\hspace{4.5cm}-{1\over2}h^{kl}\Bigl(C_{ij}+{2\over d-1}h_{ij}A\Bigr)
\\[3mm]
\dis \{\phi({\bf x},t),\pi({\bf y},t)\}={1\over2}f^2[A+\delta({\bf x}-{\bf y})]
\\[3mm]
\dis \{\phi({\bf x},t),\til{p}^{ij}({\bf y},t)\}=-{1\over2}fF^{-2}h^{ij}
[A+{1\over d}\delta({\bf x}-{\bf y})]
\\[3mm]
\dis \{\til{N}({\bf x},t),\pi({\bf y},t)\}
={1\over 2(d-1)}fNF[A+(d-1)B+\delta({\bf x}-{\bf y})]
\\[3mm]
\dis \{\til{N}({\bf x},t),\til{p}^{ij}({\bf y},t)\}
=-{1\over 2(d-1)}F^{-1}Nh^{ij}[A+(d-1)B+{1\over d}\delta({\bf x}-{\bf y})]
\\[3mm]
\dis \{\til{N}^i({\bf x},t),\pi({\bf y},t)\}={1\over2}fD^i
\\[3mm]
\dis \{\til{N}^i({\bf x},t),\til{p}^{kl}({\bf y},t)\}
=-{1\over2}F^{-2}h^{kl}D^i
\end{array}                                             \eqno(\A.1)
$$
where
$$
\left\{\begin{array}{l}
\dis A\equiv (N\Pi)^{-1}\{\Pi({\bf x},t),\del_{0}\Pi({\bf y},t)\}
-{1\over2}h^{kl}C_{kl}
\\[5mm]
\dis B\equiv N^{-2}\{N({\bf x},t),\del_{0}\Pi({\bf y},t)\}
=N^{-2}{\del(\del_{0}\Pi({\bf y},t))\over \del p_{N}({\bf x},t)}
\\[5mm]
\dis C_{ij}\equiv N^{-1}\{h_{ij}({\bf x},t),\del_{0}\Pi({\bf y},t)\}
=N^{-1}{\del(\del_{0}\Pi({\bf y},t))\over \del p^{ij}({\bf x},y)}
\\[5mm]
\dis D^i\equiv N^{-1}\{N^i({\bf x},t),\del_{0}\Pi({\bf y},t)\}
=N^{-1}{\del(\del_{0}\Pi({\bf y},t))\over \del p_{i}({\bf x},t)}
\\[5mm]
\dis F\equiv \left[{\Pi\over 2\sqrt{h}}\right]^{1/(d-1)}
\\[7mm]
f\equiv \sqrt{2d/(d-1)}.
\end{array}\right.                                                \eqno(\A.2)
$$
It is noted that both of $\{\til{N}({\bf x},t),\;\pi({\bf y},t)\}$ and
$\{\til{N}({\bf x},t),\;\til{p}^{ij}({\bf y},t)\}$ do not vanish 
simultaneously for any choice of $A$ and $B$, which meanws that the conformal 
transformation is not a canonical one.
Conversely, it would be possible to map some kind of deformed PBs to 
canonical PBs of $f(R)$-type gravity.
Similar situation is that noncommutative spacetime leads to the unimodular 
gravity.\cite{CK}\\

\noi
{\large\bf A2. Examples of deformed Poisson brackets}\\[3mm]
We present examples of deformed PBs under some simplifying assumptions.
First we assume
$$
B=0.                                                              \eqno(\A.3)
$$
This assumption seems to be natural, since $p_{N}$ is constraines to be 
vanishing and appears nowhere other than $\del_{0}\Pi$.
Then we assume
$$
C_{ij}=h_{ij}C.                                                   \eqno(\A.4)
$$
This assumption is also seems natural when we consider the transformation 
properties.
Furthermore, we assume similarly to (A.3)
$$
D^i=0.                                                            \eqno(\A.5)
$$
Now we make two kinds of simplifying assumptions for $A$.
One of them is
$$
A+\delta({\bf x}-{\bf y})=0.                                    \eqno(\A.6\a)
$$
Then we have the following PB:
$$
\{\til{h}_{ij}({\bf x},t),\;\til{p}^{kl}({\bf y},t)\}
=-{1\over 2}(\delta_{i}^k\delta_{j}^l+\delta_{i}^l\delta_{j}^k)
\delta({\bf x}-{\bf y})-h_{ij}h^{kl}\left[{1\over2}C-{2\over d}\delta
({\bf x}-{\bf y})\right]                                        \eqno(\A.7\a)
$$
Here we make a further simplifying assumption that, on the right hand side, 
only the first term remains.
Then we have
$$
C={4\over d}\delta({\bf x}-{\bf y}).                             \eqno(\A.8\a)
$$
Under these assumptions, Eqs.(A.1) reduce to the following:
$$
\begin{array}{l}
\dis \{\til{h}_{ij}({\bf x},t),\pi({\bf y},t)\}
={2\over d}fF^2h_{ij}\delta({\bf x}-{\bf y})
\\[3mm]
\dis \{\til{h}_{ij}({\bf x},t),\til{p}^{kl}({\bf y},t)\}
=-{1\over2}(\delta_{i}^k\delta_{j}^l+\delta_{i}^l\delta_{j}^k)
\delta({\bf x}-{\bf y})
\\[3mm]
\dis \{\phi({\bf x},t),\pi({\bf y},t)\}=0
\\[3mm]
\dis \{\phi({\bf x},t),\til{p}^{ij}({\bf y},t)\}=f^{-1}F^{-2}h^{ij}
\delta({\bf x}-{\bf y})
\\[3mm]
\dis \{\til{N}({\bf x},t),\pi({\bf y},t)\}=0
\\[3mm]
\dis \{\til{N}({\bf x},t),\til{p}^{ij}({\bf y},t)\}
={1\over 2d}F^{-1}Nh^{ij}\delta({\bf x}-{\bf y})
\\[3mm]
\{\til{N}^i({\bf x},t),\pi({\bf y},t)\}=0
\\[3mm]
\{\til{N}^i({\bf x},t),\til{p}^{kl}({\bf y},t)\}=0.
\end{array}                                                     \eqno(\A.9\a)
$$
The other assumption for $A$ is
$$
A+{1\over d}\delta({\bf x}-{\bf y})=0.                           \eqno(\A.6\b)
$$
In this case, instead of Eq.(A.7a), we have
$$
\{\til{h}_{ij}({\bf x},t),\;\til{p}^{kl}({\bf y},t)\}
=-{1\over 2}(\delta_{i}^k\delta_{j}^l+\delta_{i}^l\delta_{j}^k)
\delta({\bf x}-{\bf y})-h_{ij}h^{kl}\Bigl[{1\over 2}C-{1\over d}\delta
({\bf x}-{\bf y})\Bigr].                                        \eqno(\A.7\b)
$$
Then we have
$$
C={2\over d}\delta({\bf x}-{\bf y}).                             \eqno(\A.8\b)
$$
Then Eqs.(A.1) reduce to the following:
$$
\begin{array}{l}
\dis \{\til{h}_{ij}({\bf x},t),\pi({\bf y},t)\}
={2\over d}fF^2h_{ij}\delta({\bf x}-{\bf y})
\\[3mm]
\dis \{\til{h}_{ij}({\bf x},t),\til{p}^{kl}({\bf y},t)\}
=-{1\over2}(\delta_{i}^k\delta_{j}^l+\delta_{i}^l\delta_{j}^k)
\delta({\bf x}-{\bf y})
\\[3mm]
\dis \{\phi({\bf x},t),\pi({\bf y},t)\}=\delta({\bf x}-{\bf y})
\\[3mm]
\dis \{\phi({\bf x},t),\til{p}^{ij}({\bf y},t)\}=0
\\[3mm]
\dis \{\til{N}({\bf x},t),\pi({\bf y},t)\}
={1\over 2d}fNF\delta({\bf x}-{\bf y})
\\[3mm]
\dis \{\til{N}({\bf x},t),\til{p}^{ij}({\bf y},t)\}=0
\\[3mm]
\{\til{N}^i({\bf x},t),\pi({\bf y},t)\}=0
\\[3mm]
\{\til{N}^i({\bf x},t),\til{p}^{kl}({\bf y},t)\}=0.
\end{array}                                                     \eqno(\A.9\b)
$$
In this case, the noncanonical transformation
$$
\bar{h}_{ij}\equiv\til{p}^{ij},\ \ 
\bar{p}^{ij}\equiv\til{h}_{ij}                                    \eqno(\A.10)
$$
makes the PBs simpler, i.e. nonvanishing PBs take the following forms
$$
\begin{array}{l}
\dis \{\bar{p}^{ij}({\bf x},t),\pi({\bf y},t)\}
={2\over d}fF^2h_{ij}\delta({\bf x}-{\bf y})
\\[3mm]
\dis \{\bar{h}_{ij}({\bf x},t),\bar{p}^{kl}({\bf y},t)\}
={1\over2}(\delta_{i}^k\delta_{j}^l+\delta_{i}^l\delta_{j}^k)
\delta({\bf x}-{\bf y})
\\[3mm]
\dis \{\phi({\bf x},t),\pi({\bf y},t)\}=\delta({\bf x}-{\bf y})
\\[3mm]
\dis \{\til{N}({\bf x},t),\pi({\bf y},t)\}
={1\over 2d}fNF\delta({\bf x}-{\bf y}).
\end{array}                                                     \eqno(\A.11)
$$
In these variables, only two PBs are noncanonical.
So deformations seem to be small, although not minimal.

\end{document}